# Topology, Analysis, and Modulation Strategy of a Fully Controlled Modular Battery Pack with Interconnected Output Ports

N. Tashakor, J. Kacetl, D. Keshavarzi, and S. Goetz

*Abstract*— Modular battery-integrated converters or so-called dynamically reconfigurable battery packs are expanding into emerging applications, including electromobility and grid storage. Although they offer many degrees of freedom, the state-of-the-art focuses on single output systems and mostly neglects potential of such systems in generating multiple controlled outputs. This paper investigates the use of the extra degrees of freedom that a reconfigurable battery offers to generate multiple outputs for the adjustable supply of various independent loads. The technology offers the potential for functional integration, exploiting already existing transistors, and for a reduction of separate dedicated power conversion stages. Interleaved output ports, where some modules are shared among different outputs, can further reduce the necessary power electronics and cost. However, using conventional modulation techniques in particular phase-shifted carrier modulation can be challenging with shared modules among multiple ports, and different control objectives can adversely impact the overall performance. Therefore, this paper proposes a strategy to decouple the control of multiple ports to enable further simplification of required power electronics. The proposed system does not require any additional active switches for the isolated port and can operate with a wide range of output voltages. Simulation and experiments verify the developed analysis.

*Index Terms*— Modular battery, modular multilevel converter, multi-port converters, electric vehicles.

## I. INTRODUCTION

ALTHOUGH renewable energy generation as well as environmental incentives has catalyzed the expansion of electric vehicles (EV) into the market, many of the challenges remain. In the recent years, range anxiety has almost tripled the capacity of the battery packs used in modern EVs [1]. Today, an EV is powered by the hard-wired serial and parallel connection of literally hundreds of cells. In addition to the increased capacity, a trend toward higher voltage levels is observed, which results in an increasing share of serial connection in batteries with the same energy capacity. Jung discusses the advantages behind a higher-voltage battery pack (i.e., 800 V) including lower weight, better efficiency, and faster charging [2-4], while Emadi et al. compare 400 V and 800 V electric drive systems [5, 6]. However, a higher number of serial connections can also introduce various problems, including more complex monitoring and balancing sub-systems, charging from established lower-voltage chargers, as well as inverter efficiency at lower voltages [7-9].

The conventional high-power drive system consists of high-voltage batteries, a dc/dc converter that maintains the dc link of the main inverter(s). However, as many studies show, it is possible to further shift the operating point of the inverters by actively regulating the dc link voltage [10-12]. Regulating the dc link voltage improves the efficiency of the system, but most of the other problems including the balancing battery cells as well as fault tolerance persist [13, 14].

Ever falling prices of power electronics and improved performance of low-voltage transistors have led to the expansion of battery-integrated modular multilevel converters, also known as dynamically reconfigurable batteries [15-19]. These systems break the conventionally hard-wired battery pack into multiple modules with voltages often below 100 V and integrate them with power electronics to achieve dynamic reconfiguration. A modular battery system is capable of balancing, benefits from better fault tolerance, and can achieve faster output regulation [20-23]. Furthermore, it was demonstrated that the overall efficiency can be improved [24-27]. However, whereas many different topologies have been proposed for one single ac or dc load, most of these topologies ignore auxiliary loads. In conventional concepts, these would require separate sets of batteries and converters. Multiple independent power-conversion stages can further drive up the cost and also complexity of the electrical system, even though the modular converters are capable of generating all the necessary output levels with minimum extra components. Furthermore, due to the safety requirements, low-voltage auxiliary supplies are to be isolated from the high-power side, which further increases the cost and complexity of the auxiliary circuits [28, 29].

Gan et al. propose a battery-integrated cascaded half-bridge (CHB) for the battery system to provide a fault tolerant, highly modular, and fully controllable dc link voltage for driving a switched reluctance motor (SRM) [15]. However, the topology does not consider the auxiliary load requirement, and completely independent auxiliary power modules must be developed. In order to solve the problem of galvanic isolation, Kandasamy propose an inductively coupled modular battery system [30]. However, multiple high-frequency transformers as well as a high number of active components reduce the overall efficiency and increase the cost as well as volume of the system. Whereas multi-port systems are not a new concept in renewable applications, most of the available topologies are accordingly still not fit for EV applications with a modular cascaded battery system [31, 32]. This paper fills this gap by presenting a generic topology and control for both isolated and non-isolated ports based on a modular reconfigurable battery system with minimum added passive components. To develop the control strategy for the ports, first a more general case of phase-shifted carrier modulation is presented that can occur in cases where ports are connected to only a share of the available modules leading to an uneven distribution of corresponding carriers. Additionally, the gain of each port considering the interaction of modules as well as the modulation technique is derived. Finally, we propose a closed-loop strategy to decouple the control of multiple interconnected ports in the system for each case. This is all achieved by fully utilizing the available degrees of freedom offered by the modular reconfigurable battery with minimum of added components, and specifically no extra switches. The above-mentioned functions are in addition to the typical benefits of modular reconfigurable batteries, such as better output quantization, higher effective switching frequency, and improved fault tolerance.

## II. POSSIBLE TOPOLOGIES AND PORTS

Reconfigurable batteries and battery-integrated cascaded converters are not a new concept and various different macro- and micro-topologies are available in the literature [13, 14, 33-36]. Different string connections can provide dc, single-phase, and multi-phase structures with specific features [16, 17]. The string refers to a combination of one or more battery-integrated module connected to each other, where at least one output port is connected to the beginning and end of that string, as depicted in Fig. 1(a). It is possible to have only one connection between every two side-by-side modules (e.g., in case of half-bridge or bi-directional full-bridge) allowing for only bypass and series connection in the string or there can be more than one point of connection (e.g., diode-clamped [37, 38], FET-clamped [39, 40], or topologies with higher number of switches [41-43]). On the modules level, half-bridge circuits as the simplest form, can achieve unidirectional series and bypass modes (see Fig. 1(b)), while more complex electronics can offer bidirectional series, parallel, fault protection [35, 44, 45]. Fig. 1(b) illustrates the two simplest module topologies with a minimum number of semiconductors [41].

The resulting output voltage at the port terminals is the sum of the series-connected battery voltages considering their connection mode at each instant per

$$V_p^t = \mathbf{S}^{t(T)} \times V_M^t, \quad (1)$$

where $V_p^t$ is the voltage at port terminals at time $t$, $\mathbf{S}^t$ is the vector of module connections, and $V_M^t$ is the vector of battery voltages. Depending on the connection of the $i^{th}$ module, $s_i \in \{0, 1\}$, where 0 and 1 illustrate whether the battery voltage is added (series) or not (bypass or parallel), respectively. Although parallel and bypass connections have almost identical impact on the output according to (1), parallel connection can improve efficiency and balance the modules [39].

In general, two types of ports can be considered for a dc port, (*i*) non-isolated, and (*ii*) isolated. While there are many possibilities for each of these ports with various numbers of active and passive components, in most applications in particular the electromobility, the cost and complexity of the system is of crucial importance. Hence, in the proposed system, the only fully controllable components for each port are considered to be the modular batteries, and no extra switch can be added.

### A. Non-Isolated Ports

The terminal of a non-isolated port is generally connected through a low-pass $LC$ filter that forms a buck converter to discharge the batteries and a boost converter when charging them. Fig. 2(a) shows equivalent representation of one string of modular batteries, where $V_{\text{base}}$ is the minimum voltage of the string at each operating point, while $V_{\text{pulse}}$ is a pulsating voltage. Fig. 2(b) depicts a possible representation of a non-isolated port that can be connected to it, where $L_1$ and $C_{\text{dc1}}$ form

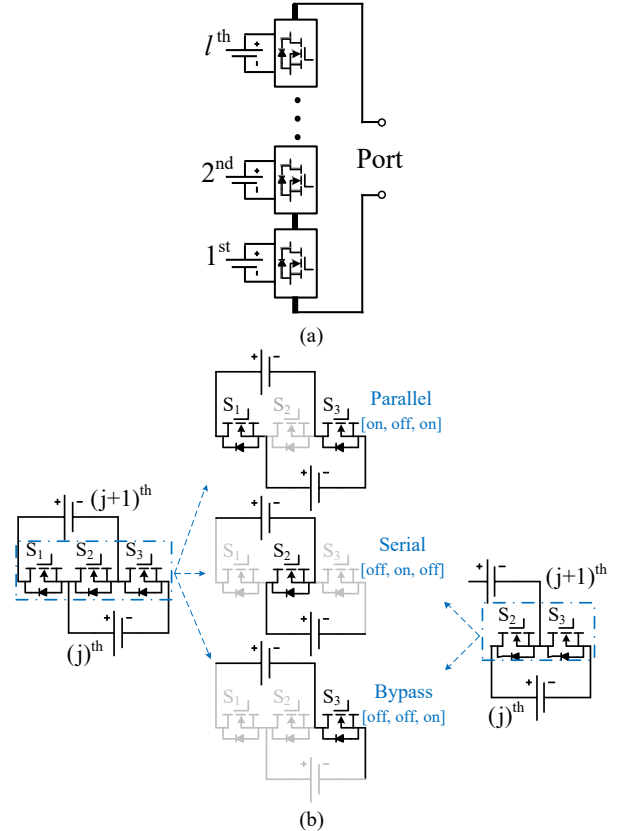

Fig. 1. (a) Generic structure of a string; (b) simplest module topologies with possible operation modes

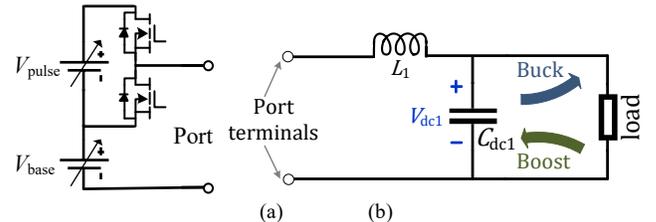

Fig.2. Circuit diagram of a non-iso port: (a) equivalent electrical circuit of the modular battery, (b) circuit of a non-isolated port

a low-pass filter.

*B. Isolated Ports*

Fig. 3(a) illustrates the circuit diagram of the envisioned isolated port. Capacitor $C_{dc2}$ and the high-frequency transformer (HFT) decouple the high-voltage string from the sensitive load through galvanic isolation. The diode-bridge provides a simple and cheap way to rectify the high-frequency voltage, and the capacitor $C_{dc3}$ provides energy to the load when the diode-bridge is not conducting [28, 46]. Fig. 3(b) represents the equivalent electrical circuit of the string with the isolated load with reference to the secondary-side of the transformer.

The next section will illustrate that depending on the number of modules and the modulation index, the amplitude and duration of the positive and negative pulses applied to the HFT can significantly vary. Having a full-bridge diode set instead of a half-bridge one can increase the freedom to control the operation point of the ports connected to the system, without actually adding a fully controlled switch. In most applications with multiple isolated and non-isolated loads, the isolated loads are for control and monitoring services, which usually have considerably lower voltage and power. Therefore, the increased cost of two extra low-power diodes is insignificant.

### III. SYMMETRIC AND ASYMMETRIC MODULATION

Modulation is the act of determining the connection modes of each module in a string, which divides into level-based such as nearest level modulation (mostly in high-voltage dc applications) and carrier-based ones such as phase-shifted carrier (PSC) modulation. PSC has the advantage of intrinsic scheduling, increased effective switching frequency, and better voltage quantization that makes it the preferred choice in low-voltage high-power applications. Since the main focus of this paper is electromobility, we consider PSC modulation for our analysis. In conventional PSC modulation, the complete switching cycle is divided among multiple carriers, where each carrier corresponds to a switch-set. For each module, the modulation index is compared to its corresponding carrier to generate the switching logics. When the modulation index is higher than the carrier, the module will connect in series and when modulation index is lower, the module will be bypassed or connect in parallel to another module [47, 48].

Normally, it is preferable to distribute the carriers as evenly as possible, to increase the symmetricity in the resulted voltage, increase voltage quantization, and reduce the required filter size at the port [35, 49]. However, doing so may not be possible in case of multiple interconnected ports, where not all the modules are included in some strings. With missing at least one module and its corresponding carrier, a gap is formed in the carrier distribution of that string, which results in what we call asymmetric PSC. Symmetric or asymmetric PSC modulations can change the behavior of the ports, which must be taken into account during the analysis and design of the component. Additionally, while symmetric PSC can achieve better distribution of pulses and lower filter requirements for a single port, asymmetric PSC can offer higher degrees of freedom to control the output of multiple interconnected ports.

*A. Symmetric (Conventional) PSC*

Assuming $N_C$ to be the number of all the nonidentical carriers in the system, the phase-difference ($\Delta\varphi$) between two consecutive carrier in symmetric PSC is $\Delta\varphi = k2\pi/N_C$, where $k$ can be factors of $N_C$ ($\{k \in \mathbb{N} | \exists L \in \mathbb{N} \text{ with } k \times L = N_C\}$). Fig. 4(a) shows an intuitive representation of a system with $L = \frac{N_C}{k}$ carriers, e.g., for $N_C = 9$, $L$ can be $\{1, 3, 9\}$ to achieve conventional PSC. As long as the maximum phase-difference is $k2\pi/L$, the order of the carrier is inconsequential. For example, for $L = 3$, the voltage-shapes at the port terminal with the phase-shift vectors of $\begin{bmatrix} \frac{2\pi}{3} & \frac{4\pi}{3} & 0 \end{bmatrix}$ and $\begin{bmatrix} 0 & \frac{4\pi}{3} & \frac{2\pi}{3} \end{bmatrix}$ are identical. Fig. 4(b) illustrates the effective modulation observed from the port terminals that contains evenly distributed carriers. Through symmetric PSC, the effective carrier has an amplitude of $\frac{1}{N_C}$ with frequency of $N_C f_{sw}$.

A symmetric PSC modulation results in only two voltage levels (i.e., $V_{base}$ and $V_{base} + v_M$). Fig. 4(c) depicts that the resulted output voltage contains a pulsing voltage ($V_{pulse}$) added on top of a base level ($V_{base}$) with a dc component as well as a squared ac one. The dc component of the output voltage can be calculated per

$$V_{dc} = mLv_M, \quad (2)$$

where $m$ is the modulation index, $L$ is the number of carriers (also corresponding modules) connected between the two terminals of the port, and $v_M$ is the voltage of one module. Please note that, normally it is assumed the battery modules are balanced, otherwise the average of modules' voltages should be considered for (2). As Fig. 2 depicts, the value of $V_{dc}$ is the voltage applied to non-iso port ($V_{dc1} = V_{dc}$).

The value of the base voltage depends on carrier number ($L$) as well as modulation index ($m$) per

$$V_{base} = \lfloor mL \rfloor \times v_M. \quad (3)$$

where $\lfloor \rfloor$ denotes the floor function that returns the next smaller integer number.

Fig. 4(b) illustrate that the pulse width of the square waveform ($D$) is different than the modulation index value ($m$), and is calculated per

$$D = mL - \lfloor mL \rfloor. \quad (4)$$

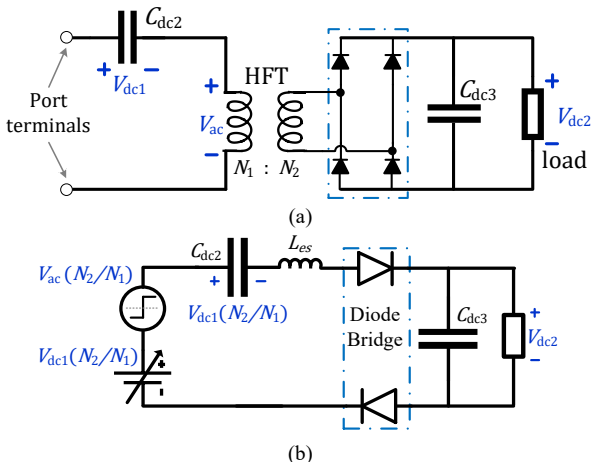

Fig. 3. Non-isolated port: (a) circuit diagram of an isolated port, (b) equivalent electrical circuit

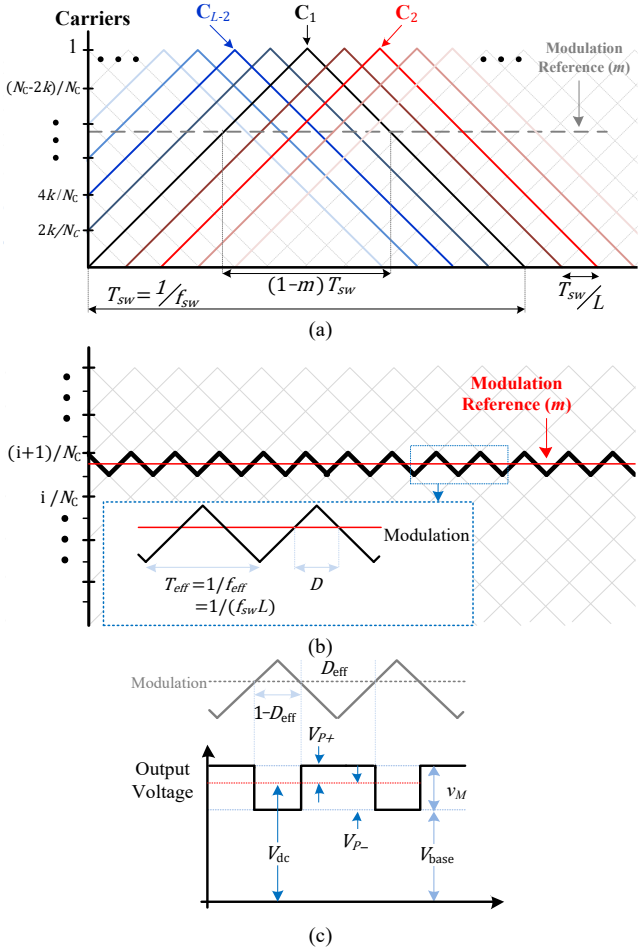

Fig. 4. Intuitive representation of PSC modulation: (a) symmetric-carriers representation; (b) equivalent representation of symmetric carriers; (C) ac and dc components of the voltage at the port terminals

Consequently, the amplitudes of positive pulse ($V_{p+}$) and negative pulse ($V_{P-}$) are

$$\begin{cases} V_{P+} = (1-D)v_M \\ V_{P-} = -Dv_M \end{cases}. \quad (5)$$

After removing the dc component of the voltage in an isolated port, the voltage can feed an isolated load through a high-frequency transformer, as depicted in . The full-bridge rectifier conducts only the voltage level with larger amplitude ($C_{dc2}$ is charged) and blocks the other voltage level ($C_{dc2}$ is discharged into load). According to (5), for $D \leq 0.5$, the positive pulse is larger ($V_{P+} \geq V_{P-}$) and vice versa. Therefore, neglecting parasitics, the average voltage at the load terminal is

$$V_{dc2} \cong \begin{cases} (1-D)v_M\left(\frac{N_2}{N_1}\right) & 0 < D \leq 0.5 \\ Dv_m\left(\frac{N_2}{N_1}\right) & 0.5 < D < 1 \end{cases}. \quad (6)$$

where $\left(\frac{N_2}{N_1}\right)$ is the transformer ratio.

Based on (6), the ratio between $V_{dc2}$ to the voltage of one module ($V_{dc2}/v_m$) in an isolated port linearly changes from $0.5 \times \left(\frac{N_2}{N_1}\right)$ to $1 \times \left(\frac{N_2}{N_1}\right)$ and it is symmetric with respect to $D = 0.5$ line, e.g., $V_{dc2}(D = 0.35) = V_{dc2}(D = 0.65)$. However, as $D$ deviates from 0.5, due to resistive elements and leakage inductances, the nonlinearities increase, and a more practical operation range of $D$ would be between 0.2 and 0.8.

Additionally, the average amplitude of diode-bridge current is

$$I_d \cong \frac{P_{load}}{V_{dc}\min[D,\ 1-D]}. \quad (7)$$

According to (7), when $D \to 0$ or $D \to 1$, the peak current goes to infinity. Therefore, duty cycles close to zero or one can result in transformer saturation or damage to the diode-bridge.

All the previous equations are under the assumption that the parasitic inductance of the high-frequency transformer is negligible. Considering the system parasitic inductance, the analysis holds for $D \leq 0.5$ as long as

$$L_{es} < \frac{D(1-D)V_{dc2}^2}{4f_{sw,eff}P_{load2}}, \quad (8)$$

where $L_{eq,p}$ is the equivalent series inductance of the system from terminals of the isolated port.

Similarly, for $D < 0.5$, the analysis is accurate as long as

$$L_{es} < \frac{(1-D)(2D-1)V_{dc2}^2}{4f_{sw,eff}P_{load2}}. \quad (9)$$

### B. Asymmetric PSC Modulation

If the carriers controlling the modulation of the modules in one string are not evenly distributed, PSC modulation will be asymmetric and the voltage at the terminals may modulate between more than two voltage levels or may not be completely uniform, as opposed to Fig. 4(c) for symmetric PSC. Such a condition can occur when modules are shared between multiple ports. Neglecting redundancies, the highest power port uses all the existing modules and their corresponding non-identical carriers ($N_C$), while the other port can share only a selection ($L$) of those modules. If $N_C$ is divisible by $L$, it is possible to have symmetric PSC for both ports, but if $N_C$ is not divisible by $L$, then at least one of the ports (usually the smaller one) will be subject to an asymmetric modulation. Fig. 5(a) illustrates the carrier's distribution in case of asymmetric PSC. In the case of asymmetric PSC, the behavior of each individual module is not changed, and the dc component of the voltage still follows (2). However, the ac component and hence the shape of voltage at the terminals of the port is considerably changed. Fig. 5(b) depicts an intuitive representation of the port voltage. Depending on $m$, $L$, and $N_C$, the shape of the voltage can vary, and it is generally asymmetric with respect to the $V_{dc}$, except for $m = 0.5$.

The output of a non-isolated port depends only on the dc component of the voltage and while some considerations are necessary to design the inductor and capacitor, the general behavior from the load side remains unchanged. However, an isolated port output depends on the maximum and minimum voltage levels.

As depicted in Fig. 5(a), if $m > (L-1)/N_C$ or $m < (N_C - L + 1)/N_C$, there are instances where all the carriers are respectively lower or higher than $m$ (see highlighted sections in Fig. 5(a)), and consequently all the modules are in series ($V_{max} = Lv_M$) or bypassed ($V_{min} = 0$). During other conditions, the maximum and minimum voltage levels can be calculated using Fig. 5(a). Therefore, the general relation of $V_{max}$ and $V_{min}$ follow

$$V_{\max} = \min\left(\underbrace{L}_{\text{Condition I}}, \underbrace{N_C - \lfloor(1-m)N_C\rfloor}_{\text{Conditions II and III}}\right)v_M, \quad (10)$$

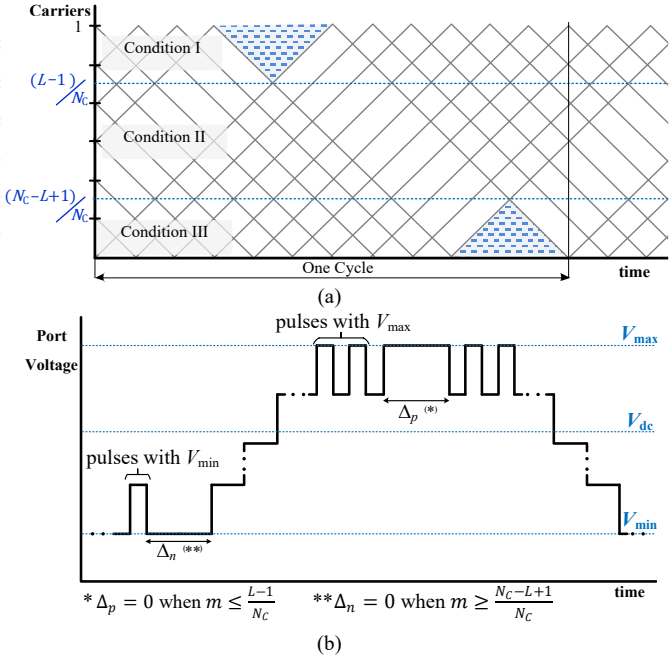

Fig. 5. Intuitive representation of asymmetric PSC modulation: (a) asymmetric representation of carriers; (b) generic voltage shape at the port terminals

$$V_{\min} = \max\left(\underbrace{L - N_C + \lfloor mN_C \rfloor}_{\text{Conditions I and II}}, \underbrace{0}_{\text{Condition III}}\right) v_M. \quad (11)$$

Knowing $V_{\max}$, $V_{\min}$, and $V_{dc}$, it is possible to calculate the amplitude of positive and negative pulses, after $C_{dc2}$ removes the dc component of the voltage per

$$V_{p+} = \begin{cases} (N_C - \lfloor (1-m)N_C \rfloor - mL)v_M, & m < L/N_C \\ (1-m)Lv_M, & m \geq L/N_C \end{cases}, \quad (12)$$

and

$$V_{p-} = \begin{cases} mLv_M, & m \leq (N_C - L)/N_C \\ (N_C - \lfloor mN_C \rfloor - (1-m)L)v_M, & m > (N_C - L)/N_C \end{cases}. \quad (13)$$

With $V_{p+}$ and $V_{p-}$, it is possible to estimate the output gain per

$$V_{dc2} \cong \max[V_{p+}, V_{p-}]\left(\frac{N_2}{N_1}\right). \quad (14)$$

The diode bridge only conducts during the largest absolute voltage level (i.e., $\max[V_{p+}, V_{p-}]$), and charges $C_{dc3}$. At other times, the capacitor ($C_{dc3}$) provides energy to the load. To calculate the current rating of the diode as well as the transformer, the effective duty cycle ($D_{eff}$) is also necessary that follows $D_{eff} = \frac{\Delta t}{T_{sw,eff}}$, where $\Delta t$ is the sum of all intervals that the diode-bridge is conducting, and $T_{sw}$ is switching cycle of one module. The total duration that the diode-bridge conducts is calculated per

$$\Delta t = \begin{cases} n_p D T_{sw,eff} + \Delta_p, & V_{p+} > V_{p-} \\ n_p D T_{sw,eff} + \Delta_p + n_n(1-D)T_{sw,eff} + \Delta_n, & V_{p+} = V_{p-}, \\ n_n(1-D)T_{sw,eff} + \Delta_n & V_{p+} < V_{p-} \end{cases} \quad (15)$$

where $n_p$ is the number of positive pulses, $D = mN_C - \lfloor mN_C \rfloor$ duty cycle of each positive pulse, and $\Delta_p$ is the interval that $m$ is above all carriers (see Fig. 5). Similarly, $n_n$ is the number of negative pulses and $\Delta_n$ is the interval that $m$ is below all carriers. Additionally, $T_{sw,eff} = T_{sw}/N_C$.

Based on the generic representation of the carriers in Fig. 5(a), the number of positive and negative pulses follows

$$n_p = \begin{cases} L - \lfloor mN_C \rfloor, & m < L/N_C \\ \lfloor m + \frac{1}{N_C} \rfloor (L-1), & m \geq L/N_C \end{cases}, \quad (16)$$

and

$$n_n = \begin{cases} -\lfloor m - \frac{1}{N_C} \rfloor (L-1), & m \leq (N_C - L)/N_C \\ L - \lfloor (1-m)N_C \rfloor, & m > (N_C - L)/N_C \end{cases}. \quad (17)$$

Using simple geometry, it is also possible to calculate $\Delta_p$ and $\Delta_n$ with respect to $m$ following

$$\Delta_p = \begin{cases} 0 & m < (L)/N_C \\ (mN_C - L + 1)T_{sw,eff} & m \geq (L)/N_C \end{cases}, \quad (18)$$

$$\Delta_n = \begin{cases} (mN_C - L + 1)T_{sw,eff} & m < (N_C - L)/N_C \\ 0 & m \geq (N_C - L)/N_C \end{cases}. \quad (19)$$

Knowing the effective duty cycle, we can estimate the average current of the diode-bridge considering the port power rating per

$$I_d > \frac{P_{\text{load2}}}{V_{dc2} D_{eff}}. \quad (20)$$

Similar to the conditions for the symmetric PSC, the provided analysis holds true when the equivalent series inductance ($L_{es}$) from the terminals of the port are negligible compared to the load resistance. Although providing the exact generic boundary conditions is not possible, the inequality

$$L_{es} < \frac{(D_{eff})^2 V_{dc2}^2}{4 f_{sw,eff} P_{\text{load2}}}(V_{p+} - \max[V_{p+} - V_m, V_{p-}]) \quad (21)$$

is a sufficient condition to check in case $V_{p+} > V_{p-}$. A similar condition can be derived for $V_{p+} < V_{p-}$ per

$$L_{es} < \frac{(D_{eff})^2 V_{dc2}^2}{4 f_{sw,eff} P_{\text{load2}}}(V_{p-} - max[V_{p-} - V_m, V_{p+}]). \quad (22)$$

In case the conditions defined by (21) and (22) are not met, no specific conclusion can be drawn except the gain will be lower, but the exact gain should be calculated through numerical methods.

IV. CONTROL STRATEGY

A. Symmetric PSC

With symmetric PSC, the gain of the isolated port in comparison to the voltage of one module is governed by the duty cycle. Substituting the duty cycle from (4) in (6) exposes a repetitive pattern for the gain with respect to the modulation index. Fig. 6 illustrates the voltage profile of both isolated and non-isolated ports with respect to $m$. The highest obtainable voltages are $N_C v_M$ in case of the non-isolated port and $v_M \frac{N_2}{N_1}$ in the case of the isolated one. Fig. 6 shows that there can be multiple voltage points for the non-isolated port per each operating point of the isolated one. Hence, it is possible to fully control the isolated output ($V_{dc2}$) and maintain $V_{dc1}$ close to its optimum point.

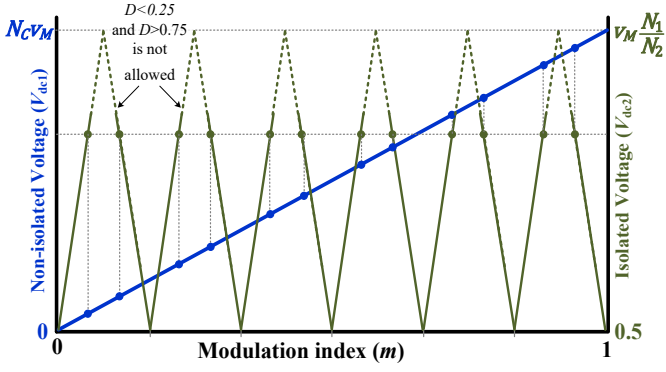

Fig. 6. Normalized output voltages with respect to $m$

According to (6), the output voltages for $D$ and $1-D$ are similar. Without the loss of generality, we will provide the analysis for $D \leq 0.5$. Additionally, the practical range of $D$ is further limited (e.g., $0.2 < D \leq 0.5$) since low values of $D$ lead to high current amplitudes and can damage the components. We can calculate $m$ per

$$m = \frac{i+0.5\pm D}{N_C}, \forall i \in \{0, 1, \ldots, N_C - 1\}, \quad (23)$$

which offers $2N_C$ operation points for $m$ to choose from, with $N_C$ being the number of available modules in the string.

Using (2), it is possible to estimate the output of each $m$ and select the point that brings the voltage of the non-isolated load closest to its optimal value. In an EV application, the non-isolated load is the dc link of an inverter–motor set, and the reconfigurable battery provides a semi-controlled dc link (normally between 400 to 800 V), while the second output is for auxiliary loads such as lighting and air conditioner (with 12 to 48 V). Therefore, the required dc link voltage for the inverter–motor set will be the optimal value for $V_{dc1}$, and the value of $m$ that brings the dc link closest to this optimal value should be selected [5, 24, 50, 51]. At the same time, the second isolated load can be fully controlled using $D$. Based on the number of modules, the maximum deviation of the dc link voltage from optimal value is $\Delta V_{\max} \leq 0.5 V_m$.

The values of module voltage and the expected voltage of the auxiliary power unit determine the transformer ratio $\left(\frac{N_2}{N_1}\right)$ following

$$\frac{V_{dc2}}{0.75 V_m} \leq \frac{N_2}{N_1} \leq \frac{V_{dc2}}{0.5 V_m}. \quad (24)$$

The generic control strategy of the system is shown in Fig. 7. The value of $V_{dc2}^{ref}$ is the rated voltage of the auxiliary power unit. $V_{dc2}$ is the measured voltage at the output of the auxiliary power unit. After calculating $D$ by the closed loop controller (e.g., simple PI controller or more complex methods), the algorithm determines the best modulation index ($m$) that brings $V_{dc1}$ closest to its reference value. At each instance, the value of $V_{dc1}^{ref}$ is provided to the system to determine the optimal modulation index ($m^{opt}$) for the dc link per (2). Next, the $m_i$ value that is closest to optimal modulation index for the non-isolated output is selected from the available choices according to (23). Hence, the output voltage of the auxiliary unit is controlled at all times, and the dc link voltage is maintained per the requirements of the main port.

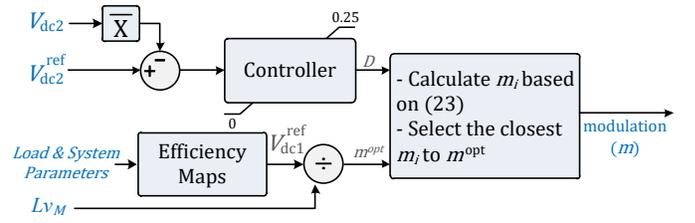

Fig. 7. The proposed control strategy for the dual-port system with symmetric PSC

### A. Asymmetric PSC

The behavior of a non-isolated load with asymmetric PSC modulation is fairly similar to the symmetric one. Although due to changed effective frequency and higher voltage variation, the passive components such as the capacitors and inductors need to be redesigned, the average dc output can be calculated using (2) and similar to the symmetric PSC, the output increases linearly by increasing $m$.

However, on the contrary to the symmetric PSC analysis for isolated ports, the profile of $V_{dc2}$ depends also on the placement of the port terminals as well as number of modules, and additionally the order of carriers. Therefore, while the provided analysis is generic, the control strategies must be developed for each specific application according to the analyzed gain profile. As an example, with nine modules ($N_C = 9$), Fig. 8 illustrates all the possible gain profiles that result in asymmetric modulation with respect to $m$. The closer the number of modules ($L$) gets to $N_C$, the closer will be the gain profile to the symmetric PSC, i.e., with higher number of choices for $m$ that lead to similar gains for the non-isolated port. As $L$ value gets closer to $N_C$, however, the range of possible gains becomes more limited. Ports with higher $L$ are more suited for loads with minimum voltage variations, to further exploit the extra operating points for other functions or to improve the control of the other output. On the other hand, as the number of modules decreases, the range of gains increases, but at the same time the instances with similar gains decrease. Therefore, the lower $L$ values are suited for loads with a wider range of output voltages.

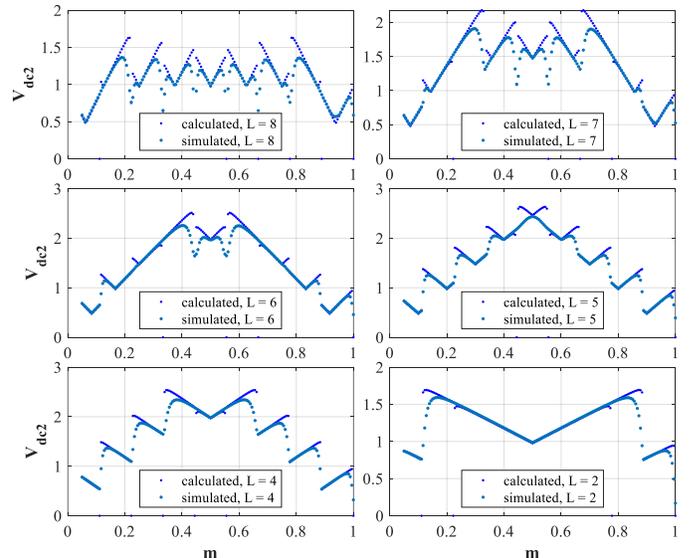

Fig. 8. Possible gain profiles with asymmetric PSC with nine-module string

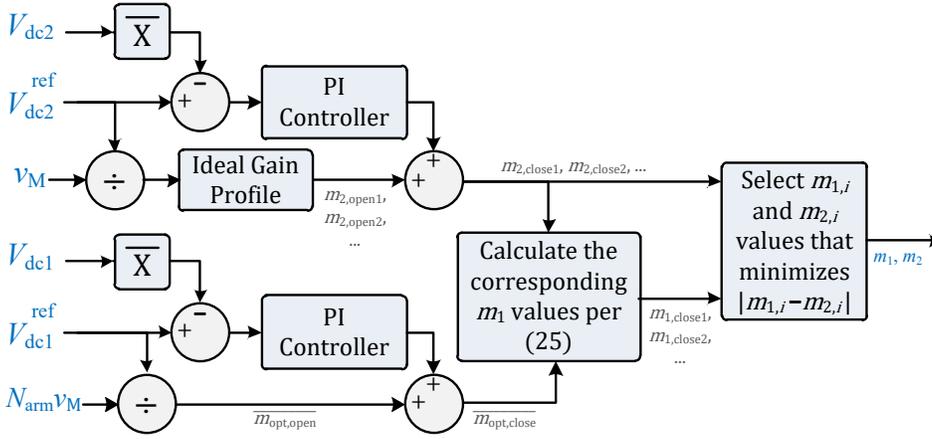

Fig. 9. The proposed control strategy for the dual-port system with asymmetric PSC

Table I
Parameters of the Simulated System

| PARAMETER | Simulation | Testbench |
|---|---|---|
| $V_{dc1}$ | $250 - 800$ [V] | $20 - 200$ [V] |
| $C_{dc1}$ | $50$ [$\mu F$] | $100$ [$\mu F$] |
| $L_{dc}$ | $220$ [$\mu H$] | $200$ [$\mu H$] |
| $C_{dc2}$ | $225$ [$\mu F$] | $940$ [$\mu F$] |
| $C_{dc3}$ | $9.6$ [$mF$] | $2.7$ [$mF$] |
| $R_{ldc}$ | $10$ [$m\Omega$] | $50$ [$m\Omega$] |
| $P_{load,1}$ | $100$ [kW] | $6$ [kW] |
| $P_{load,2}$ | $5$ [kW] | $60$ [W] |
| $R_{ds}, R_d$ | $1$ [$m\Omega$] | $1$ [$m\Omega$] |
| $V_m$ | $82 - 103$ [V] | $22 - 25.2$ [V] |
| $r_{bt,1\sim 8}$ | $1$ [$m\Omega$] | $10 - 20$ [$m\Omega$] |
| $f_{sw,eff}$ | $18$ [kHz] | $18$ [kHz] |

After identifying the gain profile according to the provided analysis in Section III, we develop the control strategy. Fig. 9 illustrates the block diagram of the proposed control strategy, which determines the modulation index of the modules that are shared between the two ports and the modules that only belong to the main (higher-power) port. If the reference value of $V_{dc2}$ varies, first the controller determines the required closed-loop gain of the second port using the ideal gain profile of the system as well as a feedback loop. Next, based on the desired closed-loop gain for the main port, controller calculates the possible $m_{2,i}$ (modulation index of the shared modules) as well as their corresponding $m_{1,i}$ (modulation index of the remaining modules). The $m_{1,i}$ are determined so that the effective equivalent modulation index is equal to $m_{opt,close}$ per

$$m_{1,i} = m_{opt,close}\left(\frac{N_C+L}{N_C}\right) - m_{2,i}\left(\frac{L}{N_C}\right) \quad (25)$$

Lastly, the best corresponding $m_1$ and $m_2$ values are chosen that minimize the modulation difference between the two groups of modules.

## V. SIMULATION AND EXPERIMENTS

### A. Simulations

MATLAB/Simulink serves for the simulation of a system with nine modules (for key parameters see Table I). In the simulation, batteries are modelled as constant dc source with resistance. Normally, the required power from the auxiliary load (few kW) in EVs is a fraction of the main load (up to hundreds of kW). We analyze the symmetric and asymmetric PSC modulation in separate scenarios. The first scenario studies a semi-controlled high-power non-isolated load (e.g., main load feeding the electric drive of an EV) and a fully controlled isolated load with fixed voltage (e.g., auxiliary circuits), both connected to all the modules resulting in symmetric PSC. The second scenario considers a varying reference voltage for the isolated load. Use of asymmetric PSC modulation for the second port, makes it possible to fully control the output voltage of the main load in a wide-range of operating conditions.

The rated voltage of each module is 96 V. Since performing an optimization is not the focus of this paper, in both cases, we assume that the optimum operating point of the main load is known and within the range of 20 % to 80 % of the maximum voltage of the battery pack, corresponding to approximately 250 V to 750 V range for an 800 V battery pack. The output voltage of the auxiliary unit in most commercial vehicles is 24 V or 48 V, where we select a fixed value of 48 V ($V_{dc2}^{ref} = 48$ V) for the first scenario and a variable voltage range of 200–400 V for the second one. Based on the module voltage range as well as $V_{dc2}^{ref}$, (25) determines the ratio of the transformer to be approximately 1.2.

*i. Scenario I: dual-port system with symmetric PSC*

In the first scenario, the system is simulated for rated battery voltages. Fig. 10 shows the voltage and current waveforms. During the simulation, the optimal reference voltage and also demand power for the first output as well as the demand power of the second output are varied. Generally, the power ratios between the non-isolated (the motor–inverter dc link) and the isolated one (auxiliaries) is roughly 2 %, which is also maintained here. The controller can provide a fixed $V_{dc2}$ under heavy variations in operation point of both outputs. Additionally, the voltage of the first output closely follows the optimal value provided as a reference to the controller. The steady-state ripple of $V_{dc1}$ and $V_{dc2}$ are below 1 % and 0.5 %, respectively. Maximum deviation of the first output from its reference value if 5%.

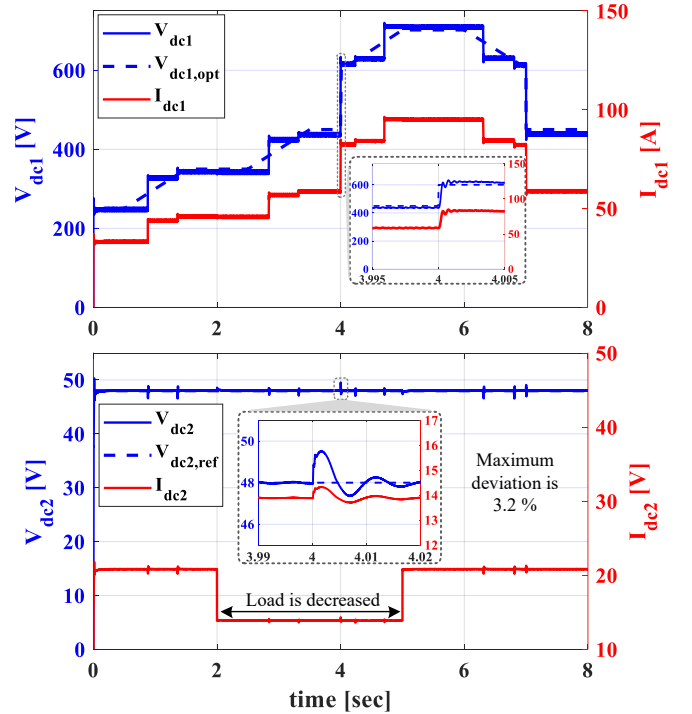

Fig. 10. Voltage and current profiles of the high-power non-isolated port as well as low-power isolated one

Fig. 11 shows the control signals as well as the voltage deviations of both outputs from their respective references. Although there are some transients at the beginning of each step, the steady-state error for the second output is less than 0.2 %. Similarly, maximum voltage deviation of $V_{dc1}$ from its optimal value is below 6 %.

    *ii.   Scenario II: dual-port system with two fully controlled outputs*

In this simulation, the auxiliary port is connected to only two of the nine modules ($L = 2$), and Fig. 8 depicts its open-loop gain profile. The procedure is similar to the first scenario, but use of asymmetric PSC enables us to fully control both outputs as proposed in Fig. 9. Both PI controllers are designed heuristically to achieve suitably damped yet fast response. The rated power of the ports remains similar to the previous scenario.

As the gain profile shown in Fig. 8 suggested, there are always two possible modulation indices for the shared modules ($m_{2,1}$, $m_{2,2}$), which are calculated based on the output of the second PI controller ($PI_2$) per

$$m_{2,1/2} = \underbrace{m_{2,open1,2}}_{from\ the\ gain\ analysis} \mp PI_2. \quad (26)$$

Correspondingly two possible $m_{1,1}$ and $m_{1,2}$ references for the remaining seven modules can be calculated based on the reference voltage of the main port using (25) and (2). Finally, the controller selects $m_1$ and $m_2$ sets that minimize the modulation difference between the modules.

As Fig. 12 shows, the output voltages of both ports closely follow the reference voltages, even under load variations at $t = 3$ sec. The maximum voltage error in both cases is $< 0.5$ %. The maximum transients are below 3 % and 4 % for the main and auxiliary outputs, respectively.

Fig. 13 depicts the calculated open-loop as well as closed-loop modulation indices for the modules, where $m_1$ is the modulation index of the modules only supplying the main port, $m_2$ is the modulation index of the two shared modules.

## B. Experiments

We built a prototype modular battery system with nine modules. The FPGA-based rapid prototyping controller (sbRIO 9726) implements the proposed control algorithms as well as the switching signals. The measurements are recorded using a LeCroy oscilloscope. Both output voltage measurements are isolated through LV25P transducers (LEM) and digitized with 16-bit accuracy to be used in the feed-back loops of the controllers. Similar to the simulations, the system consists of nine modules each with six series Li-ion cells that provide a rated voltage of 22.7 V, which results in maximum output

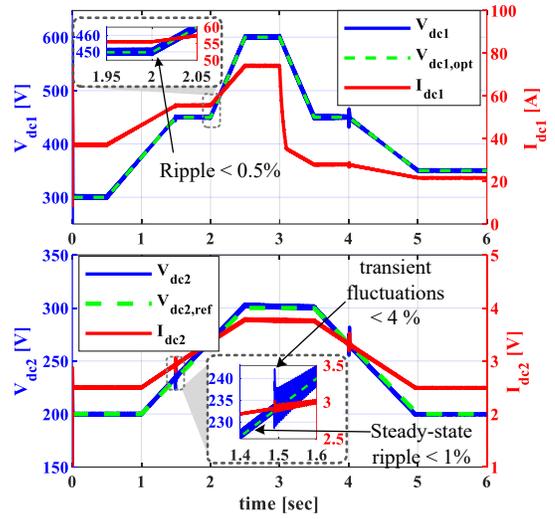
Fig. 12. Voltage and current profiles of the high-power main port as well as auxiliary one

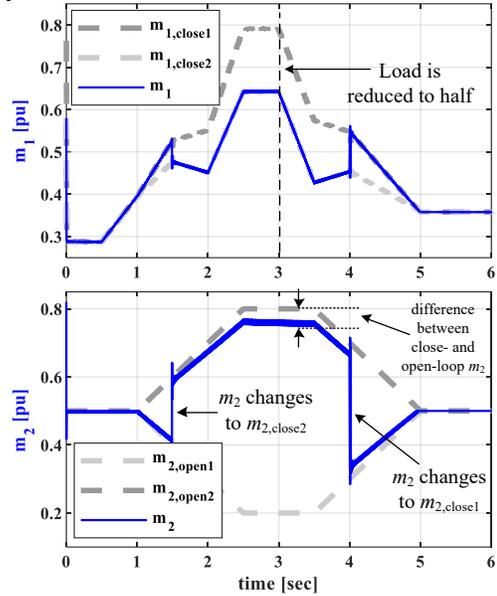
Fig. 13. The control signals as well as the outputs' deviations from reference values

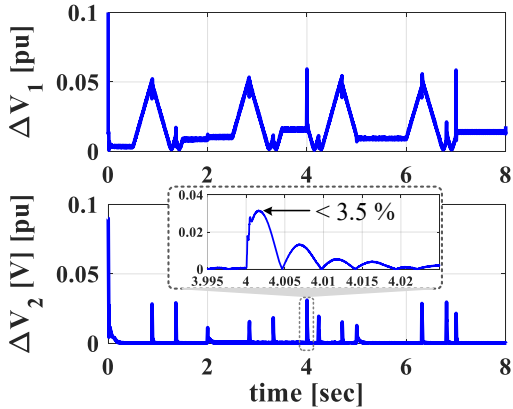
Fig. 11. The control signals as well as the outputs' deviations from reference values

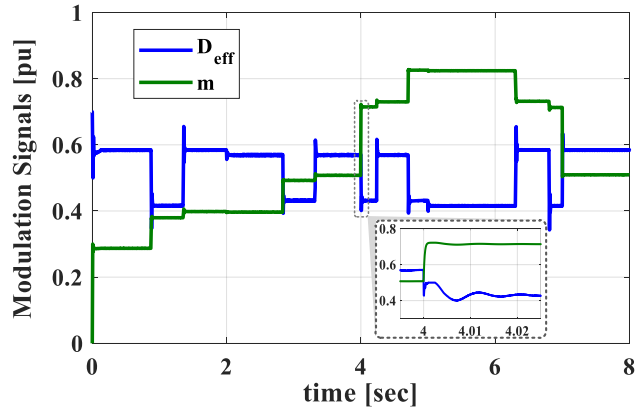

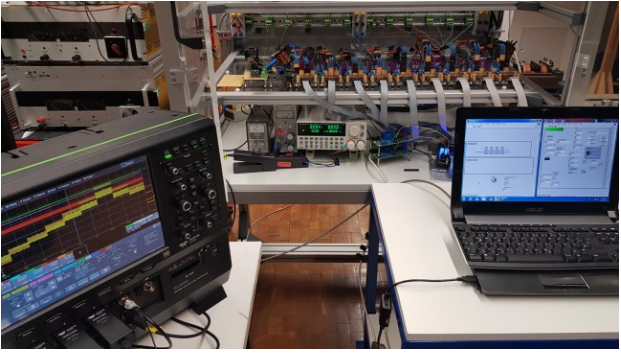

Fig. 14. Picture of the testbench

voltage of 200 V for the main output. An RL load with equivalent resistance of 10 Ω is connected to the main output, while the auxiliary load feeds an electronic load. Table I presents the parameters of the laboratory setup and Fig. 14 shows a picture of the testbench.

*i. Scenario I: dual-port system with symmetric PSC*

The desired voltage of the first dc output ($V_{dc1,ref}$) is provided as an input to the FPGA controller within the range of 12 to 198 V, while the rated output of the auxiliary port is fixed to 12 V. The controller uses the proposed algorithm to determine the most suitable modulation index and then generates the gate signals for all nine modules. Fig. 15 displays the measured voltages and currents of the main output as well as the auxiliary one with a constant load for the auxiliaries. The result confirms the performance of the controller in maintaining a constant output voltage at the auxiliary terminal when the reference voltage of the main load is changing. In such conditions, the maximum voltage ripple for the main and auxiliary outputs are below 2 %. Although using conventional PSC modulation a fully controlled output is not possible, the output voltage of the main port is maintained within a 5% of its reference value. Additionally, Fig. 16 depicts the voltage and current profile of the two outputs, when the load of the auxiliary unit is reduced to half, that confirms the fast response of the controller to load variations. The maximum transient voltage variation is 0.5 V. On the other hand, neglecting minor transients, the output voltage of the auxiliary port fully follows its reference value.

*ii. Scenario II: dual-port system with two fully controlled outputs*

The desired voltage of the first dc output ($V_{dc1,ref}$) is provided as a set point to the FPGA controller withing the range of 30 to 175 V, while the rated output of the auxiliary port changing between 100 V and 150 V. The controller uses the proposed algorithm to determine the most suitable modulation index and then generates the gate signals for all nine modules. Fig. 17 displays the measurements for output voltages and currents of the main and auxiliary ports. Using the developed controller in Fig. 9, both ports follow the output references with < 1% error. However, compared to symmetrical PSC in scenario I, the switching frequency of each module is increased to 5 kHz to achieve similar voltage ripples in both outputs.

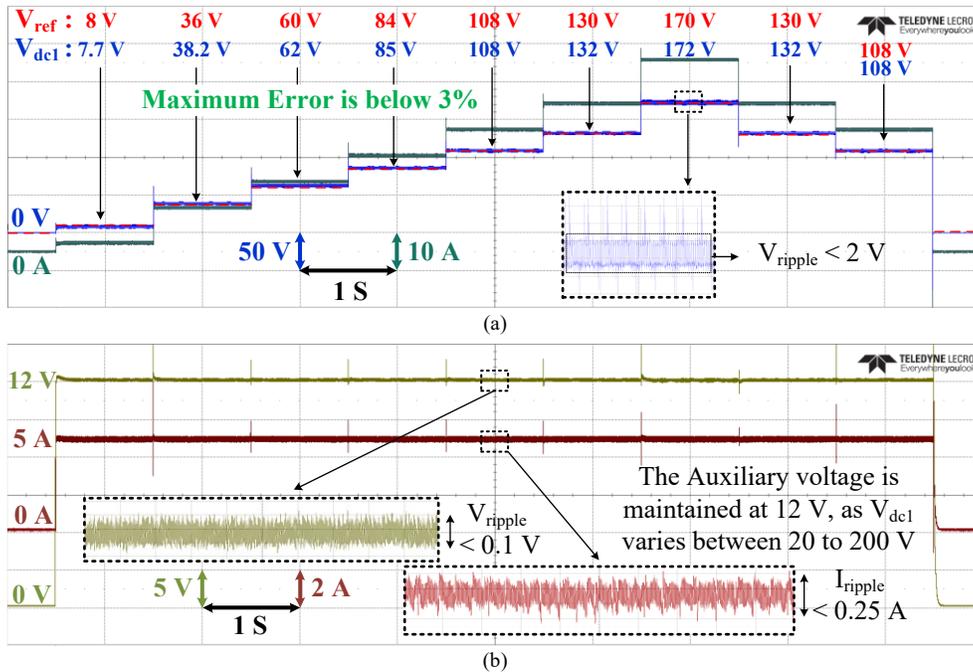

Fig. 15. Voltage and current profile of the two outputs with a constant auxiliary load: (a) main output, (b) auxiliary output

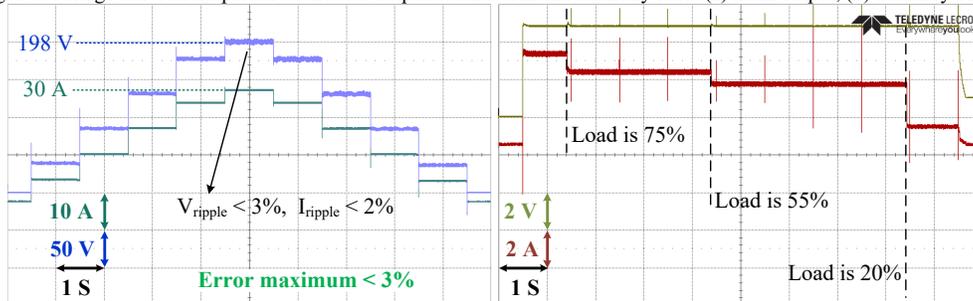

Fig. 16. Voltage and current profiles of the two outputs with load variations in auxiliaries

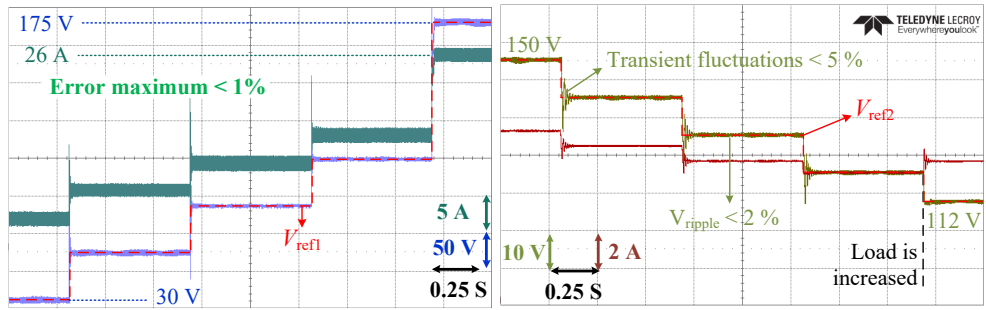

Fig. 17. Voltage and current profiles of the two outputs with load variations in auxiliaries based on asymmetric PSC modulation

## VI. Conclusion

This paper proposes a multi-port battery-integrated converter for e-mobility application, that can generate non-isolated controlled output voltages for the traction system of the electric vehicle, while using the same output voltage to generate a lower power isolated output for the auxiliary system. Additionally, the paper investigates possible interactions among interconnected output ports with conventional and/or asymmetric phase-shifted carriers as a means to increase controllability of the system. The developed strategy to decouple the control of multiple ports enables further simplification of required power electronics. The proposed system does not require any additional active switches for the isolated port and can operate with a wide range of output voltages.

The paper fully analyzes the behavior of the system under different scenarios and provides design guidelines for both the controller and topology. The provided analysis as well as simulation and real measurements support the applicability and performance of the system. The outputs can be maintained within a tight boundary of their set points with minimum added components. Furthermore, the system benefits from the typical advantages of the modular multilevel converters such as reduced filter size due to higher equivalent switching frequency, better voltage quantization, and a self-balancing capability.